\documentclass{article}
\usepackage{graphics}
\input epsf.sty
\begin{document}
\newcommand{\Abstract}[2]{{\footnotesize\begin{center}ABSTRACT\end{center}
\vspace{1mm}\par#1\par
\noindent
{~}{\it #2}}}

\newcommand{\TabCap}[2]{\begin{center}\parbox[t]{#1}{\begin{center}
 \small {\spaceskip 2pt plus 1pt minus 1pt T a b l e}
 \refstepcounter{table}\thetable \\[2mm]
 \footnotesize #2 \end{center}}\end{center}}

\newcommand{\TableSep}[2]{\begin{table}[p]\vspace{#1}
\TabCap{#2}\end{table}}

\newcommand{\FigCap}[1]{\footnotesize\par\noindent Fig.\ %
 \refstepcounter{figure}\thefigure. #1\par}

\newcommand{\TableFont}{\footnotesize}
\newcommand{\TableFontIt}{\ttit}
\newcommand{\SetTableFont}[1]{\renewcommand{\TableFont}{#1}}

\newcommand{\MakeTable}[4]{\begin{table}[htb]\TabCap{#2}{#3}
 \begin{center} \TableFont \begin{tabular}{#1} #4
 \end{tabular}\end{center}\end{table}}

\newcommand{\MakeTableSep}[4]{\begin{table}[p]\TabCap{#2}{#3}
 \begin{center} \TableFont \begin{tabular}{#1} #4
 \end{tabular}\end{center}\end{table}}

\newenvironment{references}%
{
\footnotesize \frenchspacing
\renewcommand{\thesection}{}
\renewcommand{\in}{{\rm in }}
\renewcommand{\AA}{Astron.\ Astrophys.}
\newcommand{\AAS}{Astron.~Astrophys.~Suppl.~Ser.}
\newcommand{\ApJ}{Astrophys.\ J.}
\newcommand{\ApJS}{Astrophys.\ J.~Suppl.~Ser.}
\newcommand{\ApJL}{Astrophys.\ J.~Letters}
\newcommand{\AJ}{Astron.\ J.}
\newcommand{\IBVS}{IBVS}
\newcommand{\PASP}{P.A.S.P.}
\newcommand{\Acta}{Acta Astron.}
\newcommand{\MNRAS}{MNRAS}
\renewcommand{\and}{{\rm and }}
\section{{\rm REFERENCES}}
\sloppy \hyphenpenalty10000
\begin{list}{}{\leftmargin1cm\listparindent-1cm
\itemindent\listparindent\parsep0pt\itemsep0pt}}%
{\end{list}\vspace{2mm}}

\def\TYLDA{~}
\newlength{\DW}
\settowidth{\DW}{0}
\newcommand{\dw}{\hspace{\DW}}

\newcommand{\refitem}[5]{\item[]{#1} #2
\def\REFARG{#3}\ifx\REFARG\TYLDA\else, {\it#3}\fi
\def\REFARG{#4}\ifx\REFARG\TYLDA\else, {\bf#4}\fi
\def\REFARG{#5}\ifx\REFARG\TYLDA\else, {#5}\fi.}

\newcommand{\Section}[1]{\section{#1}}
\newcommand{\Subsection}[1]{\subsection{#1}}
\newcommand{\Acknow}[1]{\par\vspace{5mm}{\bf Acknowledgements.} #1}
\newcommand{\Y}{Y_\ell^m}
\newcommand{\vxi}{\mbox{\boldmath{$\xi$}}}
\newcommand{\vx}{\mbox{\boldmath{$x$}}}
\newcommand{\vnab}{\mbox{\boldmath{$\nabla$}}}

\pagestyle{myheadings}

\def\thefootnote{\fnsymbol{footnote}}
\begin{center}
{\Large\bf Nonradial mode excitation as the cause of the Blazhko effect in RR Lyrae stars\\}
\vskip3pt {\bf W.A. Dziembowski$^{1,2}$~~ and~~
T. Mizerski$^1$} \vskip3mm {$^1$ Warsaw University Observatory,
Al.~Ujazdowskie~4, 00-478~Warszawa, Poland\\ e-mail:
wd,tm@astrouw.edu.pl\\ $^2$ Copernicus Astronomical Center,
Warsaw, Poland} \vskip5mm
\end{center}

\Abstract{A significant fraction of RR Lyrae stars exhibits
amplitude and/or phase modulation known as the the Blazhko effect. The
oscillation spectra suggest that, at least in most of the cases,
excitation of nonradial modes in addition to the dominant radial modes
is responsible for the effect. Though
model calculations predict that nonradial modes may be excited, there are 
problems with explaining their observed properties in
terms of finite amplitude development of the linear instability. We propose a
scenario, which like some previous, postulates energy transfer
from radial to nonradial modes, but avoids those problems. The
scenario predicts lower amplitudes in Blazhko stars. 
We check this prediction with a new analysis of the
Galactic bulge RR Lyrae stars from OGLE-II database. The effect is
seen, but the amplitude reduction is smaller than predicted.}

\Section{Introduction}
Explaining periodic, or nearly periodic, long term modulation of the light
curve seen in many RR Lyraes - known as  Blazhko effect -
constitutes perhaps the greatest challenge to stellar pulsation
theory. We cannot claim that we understand these most classical
pulsating stars until this effect is explained.
The idea that excitation of nonradial mode may be
responsible for the effect has been vaguely suggested long time ago (Stellingwerf,
1976). More specific suggestions regarding excitation mechanism
came later (Cox, 1993; Kov\'acs; 1993;  Van Hoolst \&
Waelkens; 1995) and these were followed by a detailed stability
calculations for a realistic RR Lyrae star model carried by Van Hoolst et al.
(1998).

A more direct observational evidence that nonradial modes may be excited in RR Lyrae stars was
provided by  Olech et al.(1999), who found close peaks in oscillation spectra of certain
RRc stars. Similar results, for a large number of RRc and RRab stars, 
were obtained in subsequent investigations (Moskalik, 2000;  Alcock et
al. 2000; Olech et al. 2001; Kov\'acs 2001; Moskalik \& Porreti, 2003,
Mizerski, 2003; Soszy\'nski et al. 2003). The decisive observational confirmation that the Blazhko
effect is caused by nonradial mode excitation may come only from analysis
of line profile changes. Such an analysis has been undertaken by Kolenberg
(2002), but the result was not sufficient for a firm conclusion. More data are needed.

The term Blazhko stars is now used to denote all objects whose
oscillation spectra exhibit peaks with frequency spacing much closer than
spacing between consecutive radial modes. The following subtypes have been
introduced (e.g. Moskalik \& Porreti, 2003) : RR-$\nu1$ stars with two
closely spaced frequencies, RR-$\nu2$ stars with three nonequidistant
frequencies, and RR-BL stars with three equidistant frequencies.  The most common are RR-$\nu1$ stars. 
RR-BL stars, in the time-domain, exhibit periodic modulations in various
forms. If amplitudes of the two side peaks are equal, then depending on the
phase relation, we either see pure phase or pure amplitude modulation. If
the amplitudes are not equal, both phase and amplitude are subject to
periodic modulation. Both amplitude and phase are also modulated in RR-$\nu1$ stars.
A quasi-periodic modulation occurs in RR-$\nu2$ case if one of the three
peaks has much lower amplitude than the others.

We cannot see any alternative to the nonradial mode excitation
as the explanation of RR-$\nu1$ and RR-$\nu2$.
In the case of RR-BL, it is perhaps natural
to look first for physical interpretation of the modulation
period, rather than the small frequency spacing. In most of such interpretations so
far the period was associated with rotation, but those interpretations were
never supported by a sound observational evidence
nor by a correct model calculations. A resonant coupling between radial modes
may, in principle, lead to a periodic limit cycle (Moskalik, 1986), but
there is no required resonance in realistic RR Lyrae models.
Thus, in our work, we consider only those explanations of Blazhko effect
that invoke nonradial mode excitation.

As shown by  Van Hoolst et al. (1998) and Dziembowski \& Cassisi (1999), there are two ways
to excite such modes in RR Lyrae stars. One is through opacity driving and the other is
through parametric instability of the radial mode to excitation of nonradial modes
at nearly resonant frequency (the 1:1 resonance). In both
cases the $\ell=1$ modes, with frequencies close to radial modes, are most easily excited.
However, the latter mechanism was preferred because the
growth rate of nonradial modes is much
lower than those of radial modes, which should quickly saturate the linear driving effect.

Nonlinear development of the parametric instability was studied with the amplitude
equation formalism by Nowakowski \& Dziembowski (2001).
After adopting certain approximations, of which
the most severe was the use of adiabatic eigenfunctions, they found that the
instability leads to pulsation involving radial and nonradial mode(s) with constant amplitude.
If there is only one nonradial mode, then the nonlinear
synchronization leads to monoperiodic pulsation, so that model
does not apply to Blazhko stars. If the coupling involves a pair
of modes of opposite azimuthal orders, $m$, in a rotating star,
then synchronization enforces equal distances between the side peaks
and the central peak. Thus, this model looked like a plausible
explanation of the RR-BL stars, but the other subtypes remained
unexplained

A severe problem for any explanation of the Blazhko effect in
terms of nonradial mode excitation was discovered later by the same
authors (Nowakowski \& Dziembowski, 2003). What they found was that
at observed surface amplitude the {\it bona fidae} nonradial modes become
strongly nonlinear in deep interior, and should be damped much more than 
the linear theory implies. This finding certainly
invalidates approximations they adopted in their earlier work. 
The question is whether the whole idea of nonradial mode excitation in RR Lyrae stars can survive. This is what we are trying to see in this paper.

\Section{Consequences of strong nonlinearity of nonradial modes in deep interior}

Let us consider a low degree nonradial oscillation of frequency close to
that of the fundamental radial mode or the first overtone in an RR Lyrae star envelope. Here we
restrict our consideration to the deep part of the envelope, where the Brunt-V\"ais\"al\"a
frequency, $N$, is much larger than the oscillation frequency.
We begin with the linear case summarizing results of
earlier investigations (Van Hoolst et al., 1998; Dziembowski \& Cassisi, 1999 ).
Locally in this part of the envelope an oscillation mode is described as a superposition of
weakly nonadiabatic gravity waves propagating inward and outward, and the
radial displacement is accurately described by the following asymptotic formula
\begin{equation}
\xi_r={C_+{\rm e}^{{\rm i}\Phi}+C_-{\rm e}^{-{\rm i}\Phi}\over r^2\sqrt{\rho k_r}}\Y
{\rm e}^{-{\rm i}\omega t}.
\end{equation}
where $C_+$ and $C_-$ are constants, $\omega$ is the complex
eigenfrequency,
\begin{equation}
k_r=\frac{d\Phi}{dr}={\sqrt{l(l+1)}\over r}{N\over\omega}(1 +{\rm i}D)
\end{equation}
is the radial wave number, $D$ describes local dissipation,
$$ \Phi=\int^r dr k_r,$$
and the remaining symbols have their standard meaning.
Everything we need to know here about $D$ is that it is
negative (dissipation), that $|D|\ll1$ for low degree
modes, and that $D\propto\ell(\ell+1)$.
The frequency is given by $\omega_R=\Re(\omega)$ and the growth
rate by  $\gamma=\Im(\omega)$.

There are many unstable low degree nonradial modes for which
the inward and outward waves have nearly equal amplitudes. At $\ell=1$, all modes in the
wide frequency range, extending from below fundamental radial
mode to above second overtone, are unstable.
Spectra of nonradial modes are very dense as the total phase, $\Re(\Phi)$, is
large. There are maxima in the $\gamma(\omega_R)$ dependence near the radial mode
frequencies, reflecting the effect of a partial mode trapping in the
acoustic cavity. However, owing to higher amplitude in the interior, even the best trapped $\ell=1$ modes
have significantly lower growth rates than the $\ell=0$ modes.

With increasing $\ell$, the instability ranges
shrink in a consequence of increasing damping in the interior.
Disparity between the outward and inward wave amplitude increases.
The trapping effect is non-monotonic. It is weakest at
$\ell=2$ and this is why the $\ell=2$ modes are less likely to be seen in RR Lyrae stars.
At certain $\ell$-values, 7 or 8 in the vicinity of the
fundamental mode, and 4  or 5 in the first overtone vicinity,
strongly trapped unstable modes appear, for which the outward waves have negligible
amplitude. Virtually all the wave energy is then dissipated on the
way to the edge of the convective core, where the gravity wave is reflected.
Still the total rate of dissipation is much smaller than the rate of energy gain via
the opacity mechanism acting in the outer layers. This is due to a large
inward amplitude decrease in the evanescent zone separating the acoustic
and the gravity wave cavities. The growth rates of such modes are
large, but because of cancellation in the disc-averaged intensity variations, such
modes are not expected to reach detectable light amplitudes.

For these strongly trapped modes we may ignore the outward wave and
use the running wave relation,
\begin{equation}
{\partial\vxi\over\partial r}={\rm i}k_r\vxi.
\end{equation}
as the inner boundary condition to determine complex eigenfrequencies, $\omega$.
This condition may be applied anywhere within the envelope, as long as
$N\gg|\omega|$.
Formally, we may also calculate in the same way the eigenfrequencies for
low degree modes. However, these
cannot be true solutions of the {\it linear} nonadiabatic
oscillation problem for the whole star because the condition demanding
$C^+\approx C^-$ at the edge of the convective core cannot be fulfilled. 
Nonetheless, we will argue that solutions are applicable if
there is a strong {\it nonlinearity} deeper in the star interior.

The proper measure of gravity wave nonlinearity is
$$\zeta=|\vxi_H k_r|$$
where
\begin{equation}
\vxi_H={ik_rr\vnab_H\xi_r\over\ell(\ell+1)},
\end{equation}
denotes horizontal components of the displacement, which are
dominant in this case. If at some place  $\zeta\approx1$, we expect wave breaking
and a large energy dissipation. The largest values of $\zeta$ always occur in the hydrogen burning shell,
where $N/\omega$ approaches $10^3$ and $r/R\sim10^{-2}$.
The criterion for the strong nonlinearity adopted by Nowakowski \& Dziembowski (2003) was $Max(\zeta)=1$.
Using nonadiabatic eigenfunctions, they translated $Max(\zeta)=1$ to the surface
amplitude of radius and luminosity variations, from which they evaluated
amplitudes of the bolometric magnitude. The resulting values are mode dependent.
The highest surface amplitudes are reached by the modes best trapped in the
acoustic cavity. At $\ell=1$, such modes have somewhat higher frequencies
and at $\ell=2$ slightly lower frequencies than the nearest radial
modes.
The maximum amplitudes depend on the aspect, $i$, through the $\Y(i,0)$
factor. The values quoted by Nowakowski \& Dziembowski (2003) refer to aspect averaged amplitudes.
The maximum amplitude at the onset of the strong nonlinearity found in that work were about
5 - 10 mmag at $\ell=1$ and by an order of magnitude lower at $\ell=2$. 
Even the values at $\ell=1$ are lower than amplitudes of the secondary
peaks found in Blazhko stars. The numbers were obtained for one representative
model of an RR Lyrae star.

We repeated similar
calculations for a sequence of RR Lyrae models with $M=0.67M_\odot$ and $Z=0.001$ calculated by
Santi Cassisi (sequence S2 in Dziembowski \& Cassisi, 1999). One
sequence was enough because nonradial mode properties are largely
determined by the radial mode periods.
The calculated amplitudes, ${\cal A}_{\rm bol}$,  for $\ell=1$ are given in
Table 1. Side peak amplitudes in many Blazhko stars are often above 100 mmag.
The values in Table 1 indicate that the problem with strong
nonlinearity may be perhaps avoided in the case of first overtone
pulsators at the short period end, but certainly not in most of the cases.

\begin{table}[h]
\caption{Maximum amplitudes of $\ell=1$ modes near fundamental and first overtone radial modes at the onset of strong
nonlinearity in a sequence of RR Lyrae models }

\medskip

\begin{tabular}{lllllllll}
\hline
$\log T_{\rm eff}$ &$\log L$& $Y_c$&$P$&$\Delta\nu/\nu$&${\cal A}_{\rm bol}$&
$P$&$\Delta\nu/\nu$&${\cal A}_{\rm bol}$\\
\hline
 & &\hspace{1.2cm}$\vert$ & & FUND.&\hspace{0.8cm}$\vert$ & & OVER.&  \\
3.805& 1.700& 0.687& 0.638& 0.014&  2.2& 0.473&  0.012&  6.9\\
3.816& 1.696& 0.659& 0.579& 0.014&  3.8& 0.430&  0.011& 10.3\\
3.827& 1.692& 0.626& 0.526& 0.016&  6.6& 0.390&  0.017& 17.9\\
3.838& 1.688& 0.589& 0.479& 0.012& 11.6& 0.355&  0.010& 33.2\\
3.849& 1.684& 0.539& 0.436& 0.014& 18.8& 0.324&  0.011& 55.4\\
3.860& 1.681& 0.453& 0.401& 0.009& 25.8& 0.298&  0.012& 81.5\\
\hline
\end{tabular}

\medskip

$L$ - luminosity in the solar units, $P$ - period in days,
$\Delta\nu/\nu$- relative distance between radial mode and
the nonradial mode with highest amplitude at the onset of the strong nonlinearity, $A_{\rm bol}$ - the
bolometric amplitude in mmag.
\end{table}

Strong dissipation of the inward propagating g-wave  in the shell source justifies use of Eq. (3)
as the inner boundary condition,
which may be applied in the part of stellar envelope, where $|\omega|\ll N$
and $|\zeta|\ll 1$. This means that the linear asymptotics applies. 
In realistic models of RR Lyrae stars there is always a zone where both 
conditions are fulfilled.
The damping rates for modes calculated with the boundary condition
given in Eq.(3) are large, even for the best trapped $\ell=1$
modes. They are given in Table 2.  We see that, if there is no
external source of energy, the modes should decay on the time scale
of days.
The only available source of energy needed to maintain nonradial
oscillations are the radial modes, which draw energy from the
radiative flux until they reach amplitude at which driving is saturated.
The nonlinear reduction of the driving rate may be modeled with a
simple expression
$$\gamma=\gamma_{0}\left(1-{A^2\over A^2_s}\right),$$
where $\gamma_0$ denotes the linear driving rate, $A$ is the mode
amplitude (let it be the relative amplitude of the photospheric radius) and
$A_s$ is the saturation amplitude in the absence of other modes.
Consider now the energy budget in the situation, when in addition to an
unstable radial mode there is a nearby
strongly damped nonradial mode of degree $\ell$.
If we ignore frequency difference between the radial and nonradial
modes and the influence of the nonradial mode on the driving
rate, the condition for energy balance leads to the following
relation for the amplitudes

\begin{equation}
A_0^2\gamma_{0,0}I_0\left(1-{A^2_0\over A^2_{0,s}}\right)=A_\ell^2\gamma_{\ell,0}I_\ell
\end{equation}
where
$$I_\ell=\int d^3\vx\rho|\vxi|^2,$$
is mode inertia calculated with the standard normalization
$\xi_r(R)=\Y{\rm e}^{-{\rm i}\omega t}$.

If saturation of the radial mode is ignored as well, then the relative amplitude of
the nonradial mode at the surface is given by
\begin{equation}
{\cal R}_0=\sqrt{\gamma_{0,0}I_0\over\gamma_{\ell,0}I_\ell}.
\end{equation}
This quantity is listed in Table 2. Although the ratio refers to
displacement, it gives us also a rough assessment of the light amplitude
ratio. The relation between relative change of flux and the radial
displacement is $\ell$-independent.
There is a disc averaging factor of about 0.7 for the $\ell=1$ mode,
 but the observed phase relation between temperature and radius
in RR Lyrae stars acts in the opposite direction.
We see that radial modes may maintain $\ell=1$ modes with a
substantial fractional amplitude. We do not know how the energy is
fed to those strongly damped modes. What we know is, that it cannot be in
steady state manner because there is no frequency synchronization
in RR-$\nu1$ and RR-$\nu2$ stars. In this scenario the latter
type may occur due to excitation of an $m=\pm 1$ pair.

Inevitable consequence of energy dissipation by the nonradial mode
is radial mode amplitude reduction below the saturation level, which must be substantial if
the actual amplitude ratio, ${\cal R}$, approaches maximum values given in Table 2. We have from Eq.(5)
\begin{equation}
{A_0\over A_{0,s}}=\sqrt{1-\left({{\cal R}\over{\cal
R}_0}\right)^2.}
\end{equation}
There is a cost of feeding nonradial modes.
A testable prediction of this scenario is that Blazhko stars should
have lower amplitudes than monoperiodic RR Lyrae stars.

\begin{table}[h]
\caption{Maximum values of $\ell=1$ to $\ell=0$ mode amplitude
ratio for the fundamental and the first overtone doublets}

\medskip

\begin{tabular}{lllllllll}
\hline
$\log T_{\rm eff}$ &$P$&$\gamma_0$&$\gamma_1$&${\cal R}_0$
&$P$&$\gamma_0$&$\gamma_1$&${\cal R}_0$\\
\hline
 &\hspace{-0.4cm}$\vert$ & &\hspace{-0.5cm} FUND.& \hspace{1.2cm}$\vert$ &
& &\hspace{-0.4cm} OVER.&  \\
3.805& 0.670& 0.021&-0.101& 0.420& 0.472& 0.087&-0.106& 0.779\\
3.816& 0.579& 0.020&-0.110& 0.387& 0.430& 0.094&-0.115& 0.783\\
3.827& 0.526& 0.015&-0.123& 0.318& 0.390& 0.088&-0.143& 0.680\\
3.838& 0.479& 0.007&-0.141& 0.212& 0.355& 0.065&-0.199& 0.492\\
3.849& 0.436& 0.001&-0.157& 0.054& 0.324& 0.011&-0.264& 0.319\\
3.860& 0.401&-0.005&-0.175& --   & 0.298& 0.010&-0.332& 0.146\\
\hline
\end{tabular}

\medskip

$P$ - period in days,$\gamma_\ell$ - growth rates in d$^{-1}$
${\cal R}$- the amplitude ratio
\end{table}

\Section{Do Blazhko stars have lower radial mode amplitudes?}

To resolve this issue, and see if the theoretical predictions
can be empirically confirmed, we analyzed RR Lyrae stars from
OGLE-II database on the Galactic bulge (Wo\'zniak et al. 2002). We used RR Lyrae catalog
derived by Mizerski (2003). His sample contains about 2700 RR
Lyrae stars in total, with 550 RRab and about 120 RRc Blazhko
stars. Such numerous samples should be sufficient to detect
statistically significant effects and draw proper conclusions.

\begin{figure}[h]
{\par\centering \resizebox*{12cm}{8cm}{\includegraphics{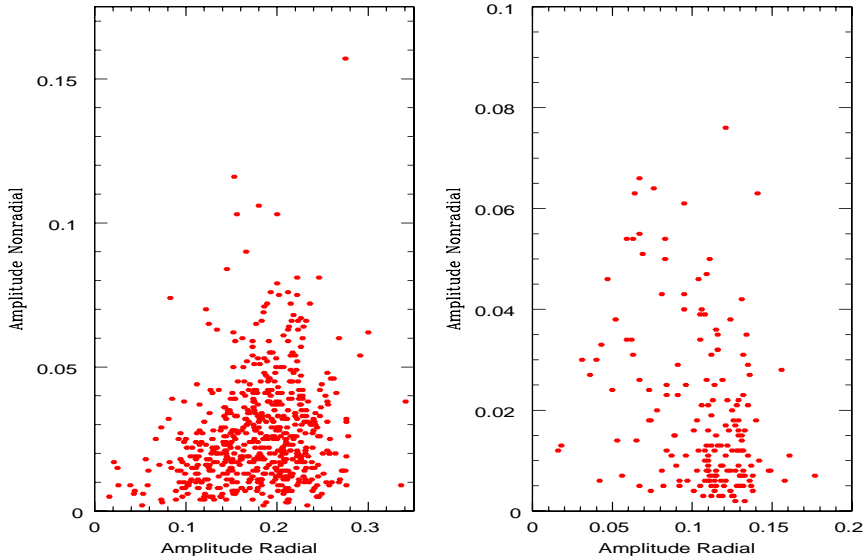}}
\par} \caption{{\small \label{radnonrad}{\em Nonradial versus
radial amplitude diagrams for RRab (left) and RRc (right) Blazhko
stars.}}\small }
\end{figure}

In the first step we analyzed the light curves and found all significant periodicities
for every star. It was done in the same way as in Mizerski (2003).
As the final result, we obtained frequencies and amplitudes of
radial pulsation modes for all RR Lyrae stars, and additionally
frequencies and amplitudes of presumably nonradial modes for all
Blazhko stars. By amplitude we mean here mathematical definition of
amplitude. It was determined as the least squares Fourier fit coefficient. 
Figure \ref{radnonrad} presents nonradial versus
radial amplitude diagrams for Blazhko stars. RRab and RRc Blazhko stars
are plotted separately.

Triangle-like shape of the diagram for RRab Blazhko stars, with
rather uniform distribution, is consistent with the interpretation
of the secondary peaks in terms of $\ell=1$ modes. The observed
amplitude for such modes should depend on the aspect angle, $i$,
through the factor $\cos i$, if the azimuthal number $m=0$. Thus,
we expect uniform random distribution of their amplitudes at
a specified radial mode amplitude. However, we also see objects
with abnormally high amplitudes that are difficult to interpret.
In the case of RRc stars the picture is less clear, perhaps
because of much smaller number of stars. When we compare the
amplitudes plotted in Fig. 1 with numbers given in Table 1 we
should remember, that numbers quoted there give amplitudes averaged
over $i$. The conclusion that the nonradial modes are nonlinear in
deep interior seems unavoidable, except perhaps for some of RRc
stars.

After identifying all Blazhko stars in our
sample, we compared the amplitudes of their radial components with
amplitudes in monoperiodic RR Lyrae stars. The amplitudes strongly depend
on the period. Therefore we binned all stars
with respect to the pulsation period. Again, RRab and RRc stars
were treated separately. As the most of RRab stars in our sample
have periods between 0.4 and 0.6 days, we only used stars from
this period range. This range was divided into four bins, each
0.05 days wide. Then, in each bin, we computed weighted average of
the radial mode amplitude. For Blazhko stars we computed two
averages: one average for all Blazhko stars and another average
for Blazhko stars with the nonradial component's amplitude larger
than 0.025 magnitude. 

\begin{figure}[h]
{\par\centering \resizebox*{12cm}{8cm}{\includegraphics{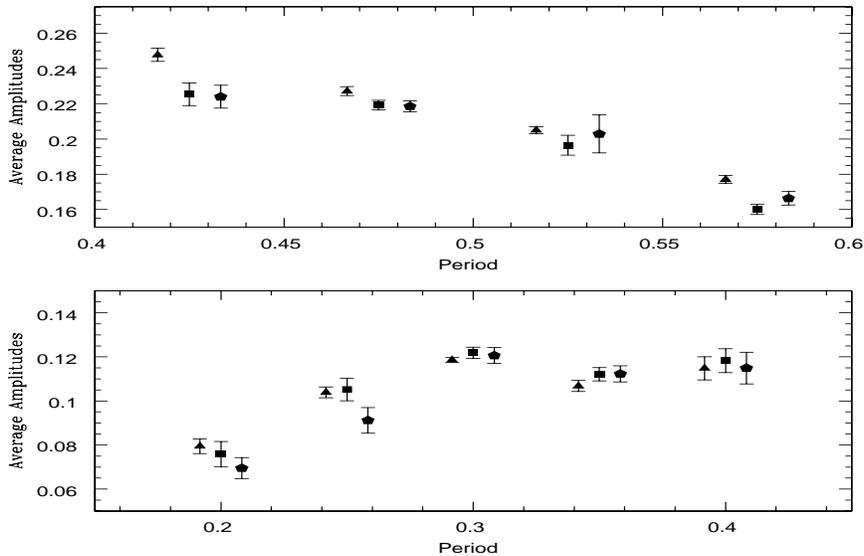}}
\par} \caption{{\small \label{averages}{\em Average amplitudes of
radial modes for RRab (top) and RRc (bottom) Blazhko stars. Triangles represent
monoperiodic RR Lyrae stars, squares represent all Blazhko stars,
pentagons represent Blazhko stars with amplitudes of nonradial
component larger than 0.025 and 0.01 magnitude for RRab and RRc
classes respectively.}}\small }
\end{figure}

The results are presented in Fig. \ref{averages}. As one
can see the presence of the Blazhko effect in RRab stars results in
lowering of the radial mode amplitude. This
effect can be seen throughout the whole period range. The stars
with higher nonradial mode amplitude do not have systematically lower
radial mode amplitude than the rest of Blazhko stars. 
This confirms that indeed most of the
difference in nonradial mode amplitudes results from the
difference in the aspect. In the case of RRc stars we do
not see any significant systematic influence of nonradial mode
presence on the radial mode amplitude.

How does the lowering of amplitudes in Blazhko RRab stars seen in
Figure 2 compare to the theoretical expectation? The typical
reduction is about 10 percent, which according  Eq.(7), corresponds
to ${\cal R}=0.45{\cal R}_0$. The values of ${\cal R}_0$ - the
maximum nonradial to radial mode amplitude ratio - for fundamental
modes are listed in Table 2. They range from 0 at the blue edge of
the instability strip to above 0.4 at the lowest temperatures.  Let us use the
0.2-0.4 range as representative for RRab stars. 
The maximum amplitude ratio, corresponding to $i=0^o$, is
by factor $\sqrt{1.5}$ higher. In Figure 1, we see stars with
${\cal R}$ exceeding the expected maximum value. These objects
certainly deserve more careful analysis. For the bulk of RRab
stars, the observed amplitude reduction appears somewhat smaller
than expected on the basis of Eq.(7) and the data in Figure 1. However, we
have to keep in mind that there is uncertainty in calculated
values of ${\cal R}_0$, following from uncertainty of $\gamma_0$.
The values of ${\cal R}_0$ for first overtone pulsations are higher
than those for fundamental mode pulsations. This may perhaps explain
why we do not see any systematic effect of the amplitude lowering
in Blazhko RRc stars.


\Section{Conclusions}

Nonradial mode excitation is the most
likely explanation of the Blazhko effect. However, there are
problems that this explanation faces. The most severe one is posed by
high amplitudes of the supposedly nonradial modes seen in many
objects. The high amplitudes imply that the modes become strongly
nonlinear and consequently damped in deep interior. We showed that
nonradial modes of spherical harmonic degree $\ell=1$ may still be
excited if there is a sufficient energy transfer from radial modes
due to nonlinear resonant coupling. The condition sets an upper
limit on the amplitude ratio. When this limit is approached, we
should see a substantial reduction of the radial mode amplitude.
We looked for this effect in observational data.

We presented a new analysis of a large sample of RR Lyrae stars
from OGLE-II database on the Galactic bulge. The Blazhko effect is 
recognized through the presence of close secondary peaks in
pulsation spectra. We emphasized that distribution of the
secondary peak amplitudes is consistent with the expected
distribution for $\ell=1$ modes due to random distribution of
aspects.

The effect of radial mode amplitude reduction is observed
only in Blazhko RRab stars, and it appears to be somewhat smaller than predicted.
There are cases, both among RRab and RRc stars, where radial and nonradial mode
amplitudes are comparable. These cases are difficult to understand
within the framework of our scenario. Possible explanation is that we
observe a special phase of a nonstationary limit cycle.

What we presented in this paper is only a schematic scenario
for maintaining nonradial oscillations.  We do not have any
detailed models of g-wave breaking in the deep interior and
nonlinear interaction between radial and nonradial modes in
outer layers. Only with such models, we will know whether we
understand how the Blazhko effect arises.

There is certainly more to be learnt from the large
sample of Blazhko stars of the Galactic bulge. More extensive analysis
of data on these objects, as well as similar data on both Magellanic
Clouds, will soon be published (Mizerski, in preparation).

{\it Acknowledgements} This work has been supported KBN grant 5
P03D 030 20.

\end{document}